\documentclass[twocolumn,showpacs,preprintnumbers,amsmath,amssymb,aps]{revtex4}
\usepackage{amsmath}
\usepackage{epsfig}
\usepackage{graphicx}% Include figure files
\usepackage{dcolumn}% Align table columns on decimal point
\usepackage{bm}% bold math

\begin{document}

\title{The excitation operator method in the spin dynamics of the one-dimensional XXZ model}

\author{Pei Wang}
\email{wangpei@zjut.edu.cn}
\affiliation{Institute of applied physics, Zhejiang University of Technology, Hangzhou, P. R. China
}

\date{\today}

\begin{abstract}
We develop the excitation operator method, which is designed to solve the Heisenberg equation of motion by constructing the excitation operators. We use it to study the spin dynamics in the one-dimensional XXZ model. We find the diffusive spin transport in the gapped phase at the high temperature limit. 
\end{abstract}


\maketitle

\section{Introduction}

Recently, the nonequilibrium dynamics of the zero and one-dimensional strongly-correlated systems attracts much attention due to the requirements of explaining a wide range of experiments in a variety of real materials. In spite of the intense efforts, the nonequilibrium 0D and 1D systems still reject a full understanding, due to the lack of a reliable numerical or analytical tool of solving the Schr\"{o}dinger equation beyond the perturbation theory. 

The attempts of tackling this problem result in a lot of new nonequilibrium methods. Most of them are based on the developments of their corresponding versions for the equilibrium systems. The examples include the real time Quantum Monte Carlo~\cite{schiro09,werner09,schmidt08,muehlbacher08}, the time-dependent numerical renormalization group~\cite{bulla08,costi97,anders05,anders06}, the scattering state numerical renormalization group~\cite{anders08}, the adaptive time-dependent density matrix renormalization group~\cite{white04,daley04,vidal03,vidal04,silva08,boulat08,meisner09,feiguin08,schollwock05}, the scattering Bethe-ansatz~\cite{mehta06}, the method based on integrability~\cite{schiller00,lesage98}, the real-time renormalization group~\cite{schoeller00,schoeller09,karrasch10,andergassen11,pletyukhov10}, the real time functional renormalization group~\cite{kennes12a,kennes12b} and the nonequilibrium flow equation~\cite{hackl07,hackl08,eckstein10,moeckel08,pei10}. However, no unique method is available which can both give the results in a high precision and incorporate the physics of the problem throughout the whole parameter space. It is still necessary to study new methods.

A usual way of driving an equilibrium system out of equilibrium is by a quantum quench, that is to switch on a Hamiltonian $\hat H$ which is not commutative with the density matrix of the system at time $t=0$. The dynamics of a physical quantity is studied by calculating the expectation value of the corresponding operator with respect to the solution of the Schr\"{o}dinger equation. Recently, the author~\cite{pei12} suggested a different way of studying the nonequilibrium dynamics by solving the Heisenberg equation of motion with the excitation operators. By decomposing the observable operator into the linear combination of a series of excitation operators, we circumvent the expansion of the S-matrix in solving the Heisenberg equation. This method can be applied in the strongly-correlated systems beyond the perturbation theory.

In this paper, we develop a modified version of the excitation operator method and use it to study the spin dynamics in the one-dimensional $XXZ$ model, which is an important paradigm of the low dimensional magnetism. The spin dynamics in a 1D system has been a field of active study for a long time~\cite{zotos05,benz05,heidarian07,langer09,langer11,karrasch12,benenti09,steinigeweg09,grossjohann10,znidaric10,lancaster10,znidaric11,prosen11,steinigeweg11,jesenko11}. There are still several open questions remaining to be settled. One of them is whether an integrable system in the gapped phase exhibits a ballistic spin transport at all temperatures~\cite{zotos05}. We contribute to this question by giving a negative answer. We find that the spin transport in the gapped phase is diffusive at the high temperature limit, contrary to the gapless phase.

The plan of the paper is the following. In Sec.~II we introduce the excitation operator method. In Sec.~III we define an orthonormal basis of the operator space in the spin-1/2 systems. The modified excitation operator method will be shown in Sec.~IV, and the results of the spin dynamics in the $XXZ$ chain will be discussed in Sec.~V. We comment on the method in Sec.~VI.

\section{Solve the Heisenberg equation of motion by constructing excitation operators}

Up to now, the analytical and numerical methods for studying the strongly-correlated system focus on constructing the quantum state, including the ground state, the density matrix at finite temperature and the time-dependent state in nonequilibrium. They reach to the target by either assuming an ansatz of the needed state or approaching gradually to it in a process of thinning out of degrees of freedom in the Hamiltonian. In these methods the fact is neglected that the quantum state contains all the information of the system, much more than what we need to calculate a physical quantity, especially a local one. Therefore, a question naturally arises: is it easier to calculate a local physical quantity than to construct the full state of the system?

In this paper we study the evolution of a physical observable in a strongly-correlated system driven out of equilibrium by solving the Heisenberg equation of motion:
\begin{eqnarray}
\frac{d\hat O(t)}{dt} = i [\hat H, \hat O],
\end{eqnarray}
where $\hat O$ is the observable operator. We solve this equation by means of constructing the excitation operators. The excitation operator of a Hamiltonian $\hat H$ is an operator $\hat A$ satisfying 
\begin{equation}\label{defexcitationop}
[\hat H, \hat A] =\lambda \hat A, 
\end{equation}
where the real number $\lambda$ denotes the excitation energy. To solve the Heisenberg equation of motion, we first decompose the observable operator into the linear combination of a series of excitation operators. Such a decomposition has been proved to always exist (see Ref.~\cite{pei12} for more details). Since the evolution of the excitation operator is trivial as $\hat A(t)= e^{i\lambda t} \hat A$, we could then get the evolution of the observable operator. At last we calculate the expectation value of the operator $\hat O(t)$ with respect to the initial quantum state at time $t=0$. To do so, we must be aware of the initial state beforehand, which is often set to be an equilibrium state and can be got by the different numerical or analytical methods designed for equilibrium systems. 

We obtain the excitation operators by the method of undetermined coefficients. We decompose the excitation operators into the linear combination of a set of operators, which form the basis of the operator space. We denote the basis operators as $\left( \hat O_1, \hat O_2,\cdots \right)$. Then we calculate the commutator between the Hamiltonian and these basis operators. The result is generally expressed in a matrix form as
\begin{equation}\label{defbasisop}
 \left[ \hat H,\hat O_i \right] = \sum_j \mathcal{H}_{j,i} \hat O_j,
\end{equation}
Now we suppose that an excitation operator $\hat A_i$ can be expressed as
\begin{equation}\label{defexinbasis}
 \hat A_i= \sum_j \mathcal{A}_{i,j} \hat O_j.
\end{equation}
Substituting Eqs.~\ref{defbasisop} and~\ref{defexinbasis} into Eq.~\ref{defexcitationop}, we obtain
\begin{equation}
 \sum_j \mathcal{H}_{i,j} \mathcal{A}_{k,j} = \lambda_k \mathcal{A}_{k,i}.
\end{equation}
So the coefficients of $\hat A$ is an eigenvector of the matrix $\mathcal{H}$ with the eigenvalue $\lambda_k$ denoting the excitation energy. These coefficients can be obtained by an arbitrary diagonalization algorithm. 

The critical step of our approach is the choice of the basis of the operator space. Since the operator space should whatever contain the observable operator, a natural way to choose the basis is to take the $\hat O$ itself as the first element of the basis. We then calculate the commutator $[\hat H,\hat O]$, which will generate new operators beyond $\hat O$. These operators are then included in the basis operators. We repeatedly calculate the commutators between the Hamiltonian and the operators in the basis, until enough number of basis operators are obtained. 

However, in the basis obtained by this way one cannot guarantee that the matrix $\mathcal{H}$ will be a real symmetric matrix, and then does not know if it is diagonalizable. This problem can be solved by constructing a set of "orthogonal" basis operators. Like what we do in a vector space, we could define an inner product between two operators, written as $ \langle \hat O_i,\hat O_j\rangle$. The inner product should satisfy all the properties of that defined between two vectors. Additionally, we set it to be always a real number. A set of basis operators are orthogonal to each other if and only if they satisfy
\begin{equation}
 \langle \hat O_i,\hat O_j \rangle = \delta_{i,j}.
\end{equation}
By using this relation and Eq.~\ref{defbasisop}, we find that the elements of $\mathcal{H}$ can be expressed as
\begin{equation}
\mathcal{H}_{i,j}= \langle \hat O_i, \left[ \hat H,\hat O_j \right] \rangle . 
\end{equation}
To guarantee $\mathcal{H}_{i,j}=\mathcal{H}_{j,i}$, we need choose a set of basis operators and an appropriate definition of the inner product, so that
\begin{equation}\label{symmrelation}
 \langle \hat O_i,[\hat H,\hat O_j ]\rangle =  \langle \hat O_j,[\hat H,\hat O_i ]\rangle .
\end{equation}
This is equivalent to say that the superoperator $[\hat H,\cdot ]$ is self-adjoint. As will be shown next, such a set of basis operators can always be found in a spin-1/2 system.

Now since $\mathcal{H}$ is a real symmetric matrix and its eigenvectors are just the rows of the orthogonal matrix $\mathcal{A}$, the basis operator $\hat O_i$ can be decomposed into the linear combination of the excitation operators:
\begin{equation}
 \hat O_i = \sum_j \mathcal{A}_{j,i} \hat A_j.
\end{equation}
Its expectation value at arbitrary time can be expressed as
\begin{equation}
  \langle \hat O_i(t) \rangle= \sum_{j,i'} \mathcal{A}_{j,i} e^{i\lambda_j t} \mathcal{ A}_{j,i'} \langle \hat O_{i'}\rangle ,
\end{equation}
where $\langle \hat O_i \rangle$ is the expectation value of the basis operator with respect to the initial state.

\section{An orthonormal basis of the operator space in spin-1/2 systems}

An orthonormal basis satisfying Eq.~\ref{symmrelation} is found in the spin-1/2 systems, where the spins are located at each site on a discrete lattice. The dimension of the operator space at each site is four. A natural choice of the onsite basis operators should include the three generators of the $SU(2)$ algebra and the identity operator. Additionally, we must guarantee the matrix $\mathcal{H}$ to be a real one. So we employ the next three operators:
\begin{eqnarray}
 \hat o^x = \left( \begin{array}{cc} 0 & 1 \\ 1 & 0\end{array}\right),  \hat o^y = \left( \begin{array}{cc} 0 & -1 \\ 1 & 0\end{array}\right),  \hat o^z = \left( \begin{array}{cc} 1 & 0 \\ 0 & -1\end{array}\right),
\end{eqnarray}
instead of the traditional $SU(2)$ generators. The commutation relations between them are $[\hat o^x,\hat o^y]=2\hat o^z$, $[\hat o^y,\hat o^z]=2\hat o^x$ and $[\hat o^z,\hat o^x]=-2\hat o^y$. An arbitrary basis operator is expressed as
\begin{equation}
\hat O = \prod_{i} \otimes \hat o_i ,
\end{equation}
where $i$ denotes the lattice site and $\hat o_i$ takes a value in the set $\{ \hat o^x, \hat o^y, \hat o^z, \mathbf{1}\}$. 

The inner product of two basis operators is defined as the trace of their product, i.e.,
\begin{equation}
 \langle \hat O_i,\hat O_j\rangle = \frac{\textbf{Tr} \left[ \hat O^\dag_i \hat O_j \right]}{2^L},
\end{equation}
where $L$ is the total number of sites in the model. It is not difficult to prove that such a definition satisfies both the positive definiteness of the inner product and the Eq.~\ref{symmrelation} as the Hamiltonian is self-adjoint. 

In this paper we focus on the one-dimensional $XXZ$ model, because it can be strictly diagonalized in the noninteracting case so that we could compare the result of our method with the exact result. The Hamiltonian of the 1D $XXZ$ model is
\begin{equation}
 \begin{split}
  \hat H = \sum_{i=1}^{L-1} \left( \hat S_i^x \hat S_{i+1}^x +\hat S_i^y \hat S_{i+1}^y +\Delta \hat S^z_i \hat S^z_{i+1} \right),
 \end{split}
\end{equation}
where the interacting strength $\Delta$ controls the magnetic order of the system at the ground state.
We are interested in the spin transport in this model. We study the evolution of the z-component of a spin located at site $m$ denoted by $\langle \hat S^z_m(t)\rangle$. For simplicity, the initial state is set to be at the high temperature limit with one spin aligned along the positive z-axis. This spin is located at site $m'$, and the $|m-m'|$ is the distance between the spin that we measure and the spin source. At the high temperature limit, the z-components of the spins are all zero except that $\langle \hat S^z_{m'}\rangle = \frac{1}{2}$, and all the high order spin correlation functions vanish. Of course, such an initial state is much simpler than the ground state of the system. But this does not mean that there is any difference when applying our method to the zero-temperature system. Since in our method we build a pure operator identity, the additional effort that one needs at zero temperature is to evaluate the expectation values of the basis operators with respect to the ground state by employing, e.g., the DMRG algorithm or the Bethe Ansatz. 

By using the basis operators we re-express the Hamiltonian as
\begin{equation}
 \begin{split}
  \hat H = \frac{1}{4} \sum_{i=1}^{L-1} \left( \hat o_i^x \hat o_{i+1}^x -\hat o_i^y \hat o_{i+1}^y +\Delta \hat o^z_i \hat o^z_{i+1} \right).
 \end{split}
\end{equation}
The observable operator $\hat o_m^z$ is called at the zeroth level. We calculate the commutator between the Hamiltonian and it and obtain
\begin{equation}
 \begin{split}
  [\hat H,\hat o_m^z]=&  \frac{1}{2}\hat o^y_m \hat o^x_{m+1} - \frac{1}{2} \hat o^x_m \hat o^y_{m+1} \\ &  +\frac{1}{2} \hat o^x_{m-1} \hat o^y_{m} - \frac{1}{2}  \hat o^y_{m-1} \hat o^x_{m}.
 \end{split}
\end{equation}
The four basis operators $\hat o^y_m \hat o^x_{m+1}$, $\hat o^x_m \hat o^y_{m+1}$, $ \hat o^x_{m-1} \hat o^y_{m}$ and $ \hat o^y_{m-1} \hat o^x_{m}$ generated above are called the operators of the first level. We then calculate the commutators between the Hamiltonian and each of them, which will generate the basis operators at the second level, and so on. We use the $L_m$ to denote the max level that we reach for the decomposition of the $\hat o^z_m$. The number of basis operators involved for the first several $L_m$ is $1,5,21,71,233,729,2223,6573,\cdots$.

\begin{figure}\label{fig:unmod}
\setlength{\unitlength}{0.0500bp}%
\begin{picture}(4500.00,3240.00)(0,0)%
\put(0,100){\includegraphics[width=0.45\textwidth]{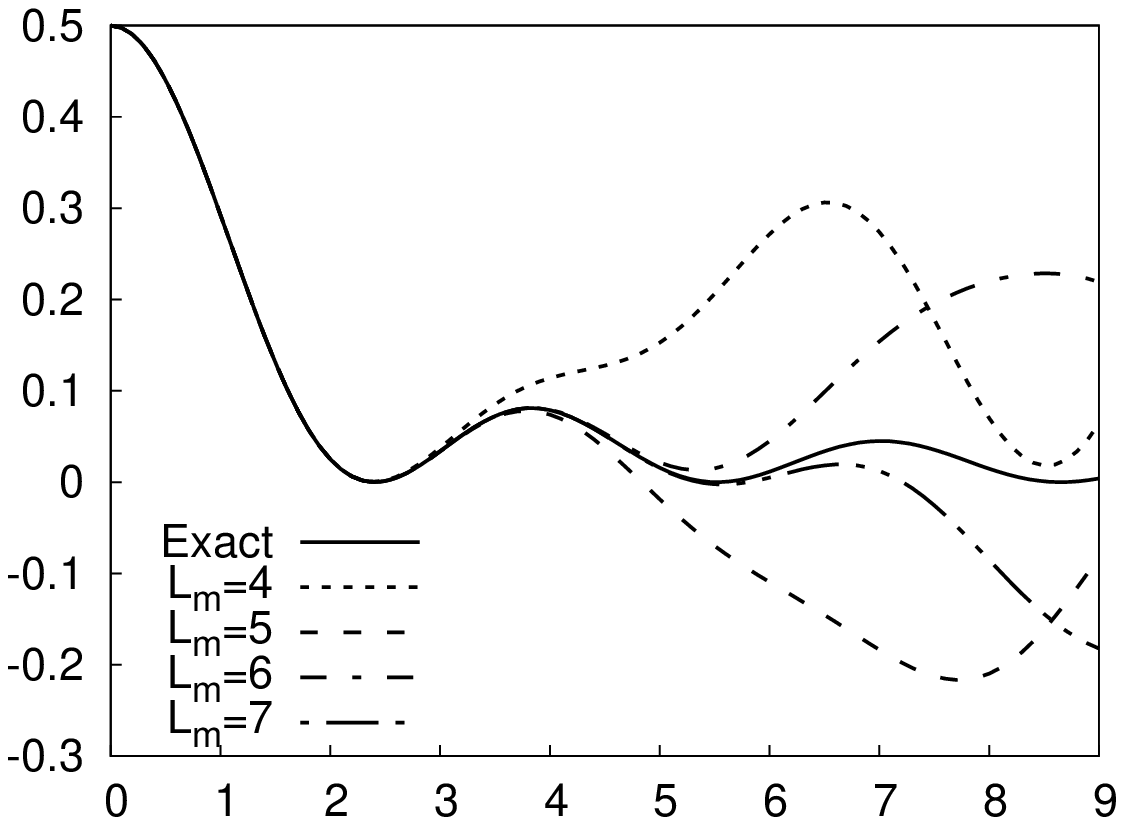}}%
\put(0,1718){\rotatebox{90}{\makebox(0,0){\strut{}$S^z_m(t) (\hbar)$}}}
\put(2540,0){\makebox(0,0){\strut{}Time}}
\end{picture}%
\caption{The spin dynamics at the site $m=m'$ as $\Delta=0$. The different max level $L_m$ is adopted in the calculation. The solid line denotes the exact result obtained by strictly diagonalizing the Hamiltonian. }
\end{figure}
Fig.~1 shows the $\langle \hat S^z_m(t) \rangle$ calculated as $L_m = 4,5,6,7$, compared with the exact result, when the interacting strength $\Delta$ and the distance $|m-m'|$ are both zero. The transient behavior of the z-component of the target spin is well predicted by decomposing it into the excitation operators. However, a strong deviation from the exact result is observed in the long-period behavior at a finite $L_m$. Increasing the $L_m$ will improve the performance at the long period, but will also increase the difficulty in diagonalizing the matrix $\mathcal{H}$. For example, as $L_m=7$ the dimension of the $\mathcal{H}$ will be $6573$! The longer the time is, the larger the dimension of the $\mathcal{H}$ should be to obtain the correct result. The error of the calculation at a finite $L_m$ can be estimated by comparing the result with that after increasing $L_m$ by one. As shown in Fig.~1, this gives the error bound. 

\section{The modified excitation operator method}

With the generation of the basis operators, the dimension of the operator space and the size of the matrix $\mathcal{H}$ will increase rapidly, and will be soon too large to the ability of the most powerful tools of diagonalization. We conquer this problem by dividing the evolution into $N$ periods, each of which has a very short time interval. The observable operator in the Heisenberg picture is then expressed as
\begin{equation}
\begin{split}
& \hat o^z_m(t) \\ & = e^{i \hat H \tau }\left( e^{i \hat H \tau }  \left( \cdots \left( e^{i \hat H \tau} \hat o^z_m e^{-i\hat H  \tau} \right) \cdots \right) e^{-i \hat H \tau }\right) e^{-i \hat H\tau},
\end{split}
\end{equation}
where $\tau= t/N$ is the time interval of a period. At each period the excitation operator method is employed to calculate the evolution of the involved basis operators $e^{i \hat H \tau} \hat O_i e^{-i\hat H  \tau}$. Since the short-time evolution can be obtained to a high precision by the excitation operator method, the $\hat o^z_m(t)$ at whatever time can be obtained explicitly as $\tau \to 0 $. By dividing the time $t$ into $N$ intervals, we both improve the performance of our method at the long time region and avoid diagonalizing a large matrix by keeping the $L_m$ at each period a small number.

However, the exponential growth of the number of involved basis operators is still kept. Since the max level that we reach in the whole evolution process is proportional to $N$, the $N$ cannot be very large. In practice, the $N$ can be greatly enhanced by neglecting the basis operators with a small weight. In fact, we find that as decomposing the $\hat O_i(\tau)$, the weights of different basis operators may vary by several magnitudes. Most of the involved basis operators has a very small weight. So we set a small number $\epsilon$ denoting the tolerance. At each period, the generated basis operators with a weight smaller than $\epsilon$ are neglected. By doing so, we could control the growth of the operator space at a low speed. And it is not difficult to estimate the error caused by a finite $\epsilon$ by letting $\epsilon\to 0$.

\begin{figure}\label{fig:delta0}
\setlength{\unitlength}{0.0500bp}%
\begin{picture}(4500.00,3240.00)(0,0)%
\put(0,100){\includegraphics[width=0.45\textwidth]{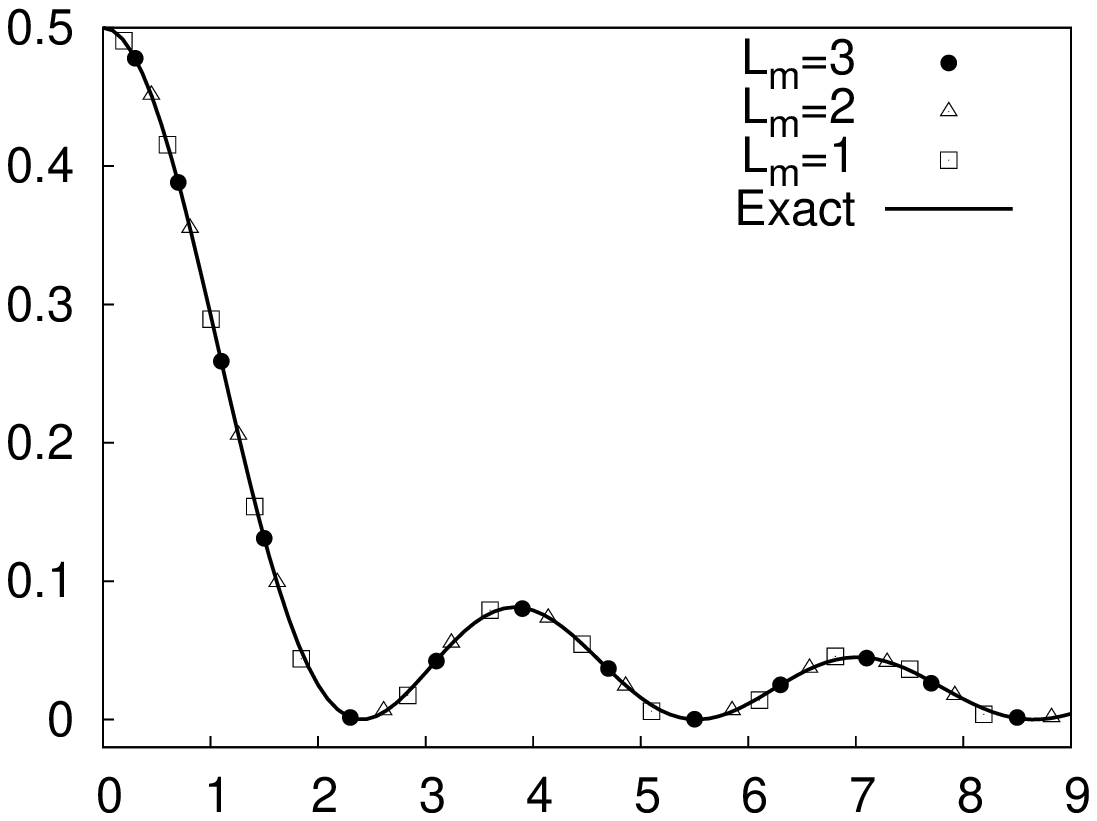}}%
\put(0,1718){\rotatebox{90}{\makebox(0,0){\strut{}$S^z_m(t) (\hbar)$}}}
\put(2540,0){\makebox(0,0){\strut{}Time}}
\end{picture}%
\caption{The simulation of the spin dynamics at $\Delta=0$ by using the modified excitation operator method. The results are compared with the exact one, denoted by the solid line. The $L_m$ denotes the max level as decomposing the basis operators in one period. }
\end{figure}
Fig.~2 shows the result of the $\langle \hat S^z_m(t)\rangle$ as $\Delta=0$. The numerical results are found to coincide well with the exact result at the time as large as $9$, when the excitation operator method without the time division fails. The exact result is obtained by strictly diagonalizing the Hamiltonian, which can be re-expressed in terms of spinless fermions $\hat c^{(\dag)}_i$ through the Jordon-Wigner transformation:
\begin{equation}
 \hat H = \frac{1}{2} \sum_{i=1}^{L-1} (\hat c^\dag_i \hat c_{i+1} + h.c.).
\end{equation}
The evolution of the density $\hat n_m(t) = \hat c^\dag_m(t) \hat c_m(t)$ can be easily worked out, so does the $\hat S^z_m(t)$. In the thermodynamic limit as $L\to \infty$ and $m=m'$, the result is
\begin{equation}
 S^z_m(t) = \frac{1}{2} J^2_{0}(t),
\end{equation}
where $J_0$ is the Bessel function of the first kind. 

Since for different $L_m$s the numerical results always converge to the exact one as the tolerance $\epsilon$ and the period $\tau$ both go to zero, how to choose $L_m$ at each period is totally decided by the efficiency of the programme. We find that keeping a smallest number of basis operators at each period, i.e., choosing $L_m=1$, will always save the computation time. The reason is that the accuracy of the result is mostly decided by the total number of basis operators involved in the calculation process. And the time cost in diagonalizing a matrix grows faster than the dimension of the matrix. So diagonalizing a lot of small matrices will be more efficient than diagonalizing a few of big ones.

One should notice that taking a large tolerance at the same time as taking a too small $\tau$ will cause a ridiculous result due to the Landau-Zener effect. Because the operator keeps almost the same when the evolution time $\tau$ is very small, and a correspondingly large $\epsilon$ will filter out all the other basis operators except itself and then the total operator space never grows. So as keeping the tolerance invariant, one should begin with a correspondingly large $\tau$. As $\tau$ goes to zero, the result at the long time region will first converge towards the exact result, and then, after the $\tau$ reaching the breakpoint, diverge as the $\tau$ decreasing more.

\begin{figure}\label{fig:dim}
\setlength{\unitlength}{0.0500bp}%
\begin{picture}(4500.00,3240.00)(0,0)%
\put(0,100){\includegraphics[width=0.45\textwidth]{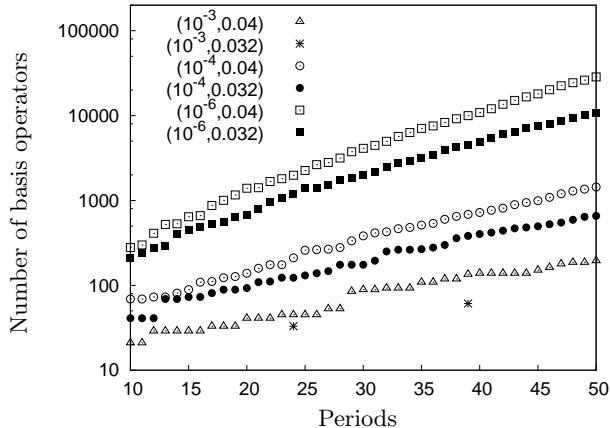}}%
\put(0,1718){\rotatebox{90}{\makebox(0,0){\strut{} Number of basis operators }}}
\put(2540,0){\makebox(0,0){\strut{}Periods}}
\end{picture}%
\caption{The growing of the number of basis operators involved at each period with the period. Each data is titled by the pair $(\epsilon,\tau)$. The interacting strength is $\Delta=0.5$.}
\end{figure}
In the modified excitation operator method, the operator space grows exponentially with the number of periods $N$ (see Fig.~3) as the $\tau$ is fixed, even after we set a non-zero tolerance. So the computational resource increases exponentially with the time, since $N$ increases linearly with the time. But if the time $t$ is fixed, increasing the $N$ will not significantly increase the computation time, because the number of involved basis operators will decrease as the time interval $\tau$ decreasing. In this case, the tolerance $\epsilon$ is critical to the computation time.

\section{The diffusive and ballistic transport in the $XXZ$ model}

The $XXZ$ model is integrable. As $\Delta \neq 0$, the ground state of the Hamiltonian has been well studied by using the Bethe Ansatz and field theoretical methods. In the case of the antiferromagnetic coupling, the system experiences a quantum phase transition at the point $\Delta=1$ from a gapless $XY$ phase to a gapped antiferromagnetic phase. However, there is no method of diagonalizing the Hamiltonian directly. To study the non-equilibrium dynamics of this system, one has to appeal to the numerical calculations.

\begin{figure}\label{fig:deltahalf}
\setlength{\unitlength}{0.0500bp}%
\begin{picture}(4500.00,3240.00)(0,0)%
\put(0,100){\includegraphics[width=0.45\textwidth]{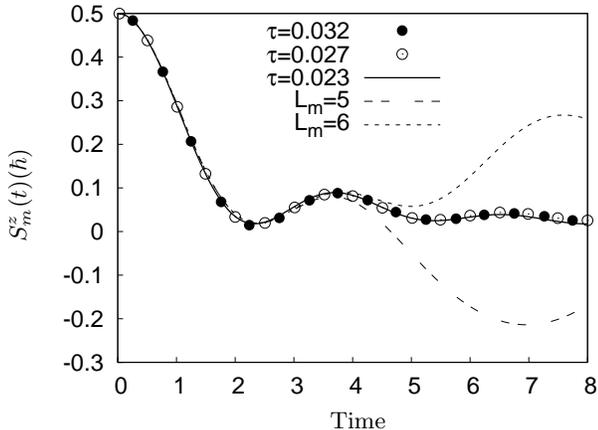}}%
\put(0,1718){\rotatebox{90}{\makebox(0,0){\strut{}$S^z_m(t) (\hbar)$}}}
\put(2540,0){\makebox(0,0){\strut{}Time}}
\end{picture}%
\caption{The spin dynamics at the site $m=m'$ as $\Delta=0.5$, calculated by the modified excitation operator method and the excitation operator method. In the calculation using the modified method, the time interval at each period $\tau$ is varied, while the tolerance $\epsilon$ is fixed to be $10^{-3}$. The curves titled $L_m=5$ and $L_m=6$ are the results got by the excitation operator method without the time division. }
\end{figure}
Fig.~4 shows the $S^z_m(t)$ as $m=m'$ and $\Delta =0.5$. The convergence of the result as $\tau$ decreasing has been observed as the tolerance $\epsilon=10^{-3}$. We also compare the result with that without the time division, in which a good estimation of the error can be obtained. We conclude that the result as $\epsilon=10^{-3}$ has been very close to the real one. Decreasing $\epsilon$ more will not result in a significant enhancement to the accuracy. And this conclusion is not changed as the distance $|m-m'|$ increasing.

\begin{figure}\label{fig:diffdelta}
\setlength{\unitlength}{0.0500bp}%
\begin{picture}(4500.00,3240.00)(0,0)%
\put(0,100){\includegraphics[width=0.45\textwidth]{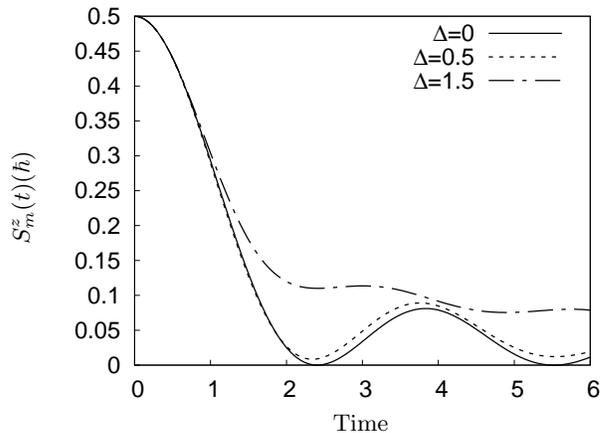}}%
\put(0,1718){\rotatebox{90}{\makebox(0,0){\strut{}$S^z_m(t) (\hbar)$}}}
\put(2540,0){\makebox(0,0){\strut{}Time}}
\end{picture}%
\caption{The comparison of the spin dynamics at the site $m=m'$ at different $\Delta$. In the calculation, the tolerance is set to be $\epsilon=10^{-3}$ and the time interval is set to be $\tau=0.04$. An intermediate quasi-steady regime is found at $\Delta=1.5$. }
\end{figure}
We study the spin relaxation at different $\Delta$ (see Fig.~5). An obvious difference is observed between the gapless $XY$ phase and the gapped anti-ferromagnetic phase: in the gapped phase there is an intermediate quasi-steady regime, which is not seen in the gapless phase. The quasi-steady $S^z_{m'}$ at the intermediate time before decaying to zero at the long time limit can be understood as the result of the antiferromagnetic ordering. A short-time local antiferromagnetic ordering around $m'$ is induced by the initial spin, which again stabilize it. 

\begin{figure}\label{fig:diffusive}
\setlength{\unitlength}{0.0500bp}%
\begin{picture}(4500.00,3240.00)(0,0)%
\put(0,100){\includegraphics[width=0.45\textwidth]{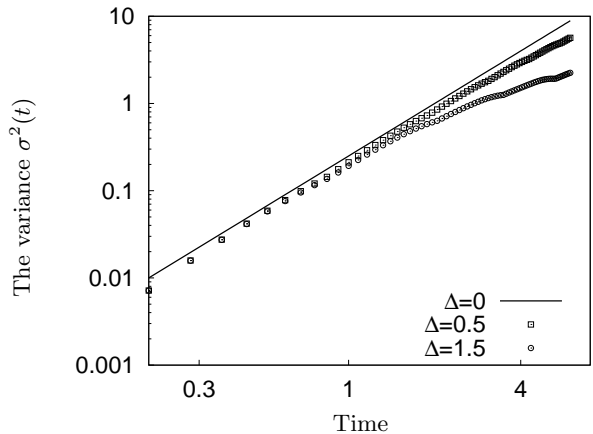}}%
\put(0,1718){\rotatebox{90}{\makebox(0,0){\strut{}The variance $\sigma^2(t)$}}}
\put(2540,0){\makebox(0,0){\strut{}Time}}
\end{picture}%
\caption{The variance functions at different $\Delta$, plotted in the logscale. The variance function is linear in the gapped phase as $\Delta=1.5$, while quadratic in the gapless phase. }
\end{figure}
The persistence of the spin in the gapped phase indicates a shift from the ballistic spin transport in the gapless phase to the diffusive transport. For quantitatively distinguishing the ballistic and diffusive spin transport, we turn to the time-dependent variance of the spin density, defined as
\begin{equation}
 \sigma^2 (t)= \sum_{m} (m-\mu)^2 \left(n_m(t)-n_m(0) \right),
\end{equation}
following the Ref.~\cite{langer09}. Here $n_m(t) =\langle \hat S^z_m (t)\rangle+\frac{1}{2}$ is the particle density function, and $\mu$ the first moment of it. In the thermodynamic limit, there is always $\mu=m'$. Especially, the variance function at $\Delta=0$ is strictly calculated to be $\sigma^2(t)= \sum_{m=1}^\infty m^2 J_m^2(t)$. In general, the variance will be a linear function of the time in a diffusive transport, while a quadratic function in a ballistic transport. We compare the variance functions at different $\Delta$. We observe the obvious character of the diffusive transport in the variance function at $\Delta=1.5$. At the same time, the variance function at $\Delta=0.5$ looks similar to the noninteracting one.

\section{Comments}

The modified excitation operator method is distinguished from most of the other real time methods by its focusing on solving the Heisenberg equation of motion instead of the Schr\"{o}dinger equation. In this aspect it is similar to the real time flow equation approach. It will result in a pure operator identity. An extra equilibrium method of calculating the initial state should be employed if the initial state is not trivial. The modified excitation operator method is suitable for a local observable, e.g., a local spin, because it save the time of getting the knowledge of the full quantum state in a high dimensional Hilbert space. It gives a very accurate result in an extremely short time in the transient regime. However, the computational resource grows exponentially as the time increasing. So it is difficult to calculate the physical quantity at the long time period by using this method. Addtionally, it is natural to perform this method in the thermodynamic limit, because the basis operators generated in course of time are all the local operators. Then the size of the system is not prerequisite to this method.

\end{document}